\documentclass{edbk} 

\usepackage{epsfig}
\normallatexbib

\begin{document}

\articletitle[AGS E917 results]      
{Results from the E917 experiment at the AGS}  

\author{Birger B. Back}
\affil{Argonne National Laboratory\\
Argonne, IL 60439, USA}
\author{for the E917 collaboration\\
 B.~B.~Back$^1$, R.~R.~Betts$^{1,6}$, H.~C.~Britt$^5$, J.~Chang$^3$,
 W.~Chang$^3$, C.~Y.~Chi$^4$,
Y.~Y.~Chu$^2$, J.~Cumming$^2$, J.~C.~Dunlop$^8$, W.~Eldredge$^3$,
S.~Y.~Fung$^3$, R.~Ganz$^{6,9}$,
E.~Garcia-Soliz$^7$, A.~Gillitzer$^{1,10}$, G.~Heintzelman$^8$,
W.~Henning$^1$, D.~J.~Hofman$^1$,
B.~Holzman$^{1,6}$, J.~H.~Kang$^{12}$, E.~J.~Kim$^{12}$, S.~Y.~Kim$^{12}$,
Y.~Kwon$^{12}$, D.~McLeod$^6$,
A.~Mignerey$^7$, M.~Moulson$^4$, V.~Nanal$^1$,  C.~A.~Ogilvie$^8$,
R.~Pak$^{11}$, A.~Ruangma$^7$,
D.~Russ$^7$, R.~Seto$^3$, J.~Stanskas$^7$, G.~S.~F.~Stephans$^8$,
H.~Wang$^3$, F.~Wolfs$^{11}$,
A.~H.~Wuosmaa$^1$, H.~Xiang$^3$, G.~Xu$^3$, H.~Yao$^8$, C.~Zou$^3$}

\affil{$^1$ Argonne National Laboratory, Argonne, IL 60439, USA\\
$^2$ Brookhaven National Laboratory, Upton, NY 11973, USA\\
$^3$ University of California at Riverside, Riverside, CA92521, USA\\
$^4$ Columbia University, Nevis Laboratories, Irvington, NY10533, USA\\
$^5$ Department of Energy, Division of Nuclear Physics, Germantown, MD
20874, USA\\
$^6$ University of Illinois at Chicago, Chicago, IL 60607, USA\\
$^7$ University of Maryland, College Park, MD 20742, USA\\
$^8$ Massachusetts Institute of Technology, Cambridge, MA 02139, USA\\
$^9$ Max Plank Institute f\"ur Physik, D-80805 M\"unchen, Germany\\
$^{10}$ Technische Universit\"at M\"unchen, D85748 Garching, Germany\\
$^{11}$ University of Rochester, Rochester, NY14627, USA\\
$^{12}$ Yonsei University, Seoul 120-749, Korea\\
}

\newpage
\begin{abstract}
Collisions of Au+Au have been studied at beam kinetic energies of 6, 8, and
10.8 GeV/nucleon at the AGS facility at Brookhaven National Laboratory.
Particles emitted from the collisions were momentum analyzed and identified
in a magnetic spectrometer system.
Measurements were made at spectrometer angles in the range 14$^\circ$ - 
59$^\circ$. $m_t$-spectra of protons from central collisions were
analyzed to derive integrated rapidity distributions and inverse slope as a 
function of rapidity. The results are compared with a thermal model and it
is concluded that there is either substantial transparency 
or longitudinal expansion at all three beam energies.
\end{abstract}


\section{Introduction}

One of the expected ways of achieving the Quark Gluon Plasma phase of
hadronic
matter is to increase the matter density to 5-10 times that of
normal nuclear matter. It has been thought that these high densities
may be achieved in central collisions of heavy nuclei at AGS
energies, provided that the relative motion is stopped. The
question of the degree of stopping in head-on collisions is thus of central
importance at these energies. Direct information pertaining to this
issue may be obtained by studying the transverse mass spectra of protons,
the majority of which are primordial, over a wide rapidity range to assess
whether the observed rapidity distribution is consistent with the initial
momentum of the projectile being converted to isotropic emission from a 
source at rest in the center-of-mass system. We find that complete stopping,
defined in this way, is not achieved for the central
Au-Au collisions at any of the energies studied in the E917 experiment.

\section{Experimental arrangement}

\begin{figure}[hb]
\centerline{
\epsfig{file=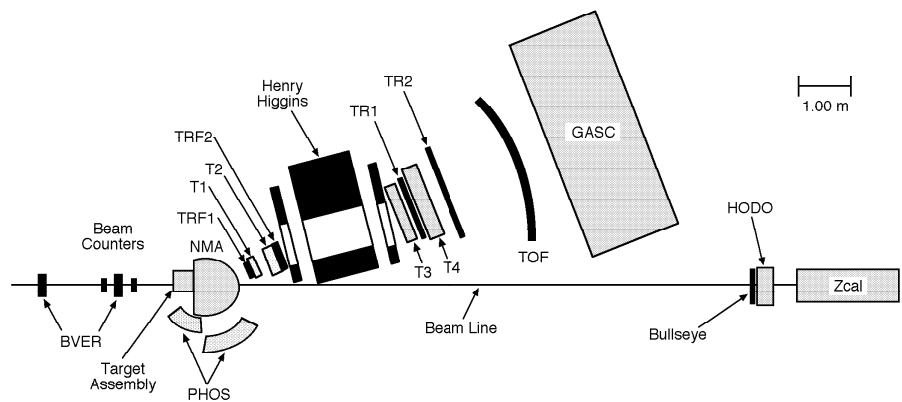,width=\textwidth}}
\caption{Experimental arrangement.}
\label{Back_fig1}
\end{figure}

The experimental arrangement is illustrated in Fig. 
\ref{Back_fig1}. Beams of $^{197}$Au
with momenta of 6.84, 8.86 and 11.69 GeV/c per nucleon corresponding to
kinetic energies of 6.0, 8.0 and 10.8 GeV per nucleon were obtained from the
AGS  at Brookhaven National Laboratory and focused onto a
Au-target of 1 mm thickness, which corresponds to $\sim$~3$\%$ interaction
probability in the target. The trajectory of each beam particle was
determined by a beam vertex detector consisting of four planes of
scintillating fibers read out by position sensitive
photo-multiplier tubes arranged in orthogonal pairs and located at 5.84 m
and
1.72 m upstream from the target position. Further beam characterization was
performed using beam-time-zero and halo counters also placed upstream from
the target. Triggers for beam interactions with the target were obtained 
by requiring that a signal of less than 75\% of that expected for the 
full energy loss of a 
Au-beam nucleus was registered in a circular ``bulls-eye'' \v{C}erenkov
detector placed 11 m downstream from the target. 

The centrality of each beam-target interaction 
was derived from the multiplicity of 
particles (mostly pions) registered in a multiplicity detector array
subtending a solid angle of about  6.85 sr around the target and/or by the 
total energy of the projectile remnant measured in the zero degree calorimeter.

In order to determine the reaction plane orientation in peripheral collisions
a hodoscope consisting of two orthogonal planes of 1 cm wide plastic
scintillator slats was placed in front of the zero degree
calorimeter. The azimuthal angle of the reaction plane determined from 
the average position of the charged 
projectile remnants in the hodoscope, relative to the beam
axis will be used to study collective flow characteristics of peripheral
collisions and the reaction plane dependence of the apparent source size
obtained in a Hanbury-Brown Twiss analysis of pion 
pairs. Such an analysis is the subject of B. Holzman's talk at this 
workshop \cite{Holzman}. 

Particle spectra were obtained by momentum analysis in a movable 
magnetic spectrometer consisting of a 0.4-Tesla magnet (Henry Higgins) and a
number of multi-wire ionization chambers used to determine the straight line
trajectories of particles entering and exiting the magnetic field. In
addition, a plastic scintillator wall located behind the spectrometer provided
particle identification by time-of-flight measurements relative to the 
\v{C}erenkov 
start detector located in front of the target.

This arrangement is capable of identifying charged pions, kaons, protons,
anti-protons and heavier nuclei. The separation of pions and kaons is
effective up to an energy of $\sim$1.75 GeV. Protons are separated from
pions up to $\sim$3.4 GeV, although there is a negligible kaon
contamination above $\sim$2.9 GeV.

The acceptance of the spectrometer is also sufficient to detect correlated
decay products of $\Phi \rightarrow K^+K^-$, $\Lambda \rightarrow p\pi^-$,
and $\overline \Lambda  \rightarrow \overline p\pi^+$. The analysis of
$\Phi$-mesons and $\overline \Lambda$ is the subject of W.-C. Chang's talk 
at this workshop\cite{Chang}.

The present talk will concentrate on the proton spectra obtained at the
three beam kinetic energies of 6, 8, and 10.8 GeV/nucleon, and an analysis
to the resulting rapidity distributions and fitted inverse slopes obtained
from the measured $m_t$-spectra.

\begin{figure}[ht]
\centerline{
\epsfig{file=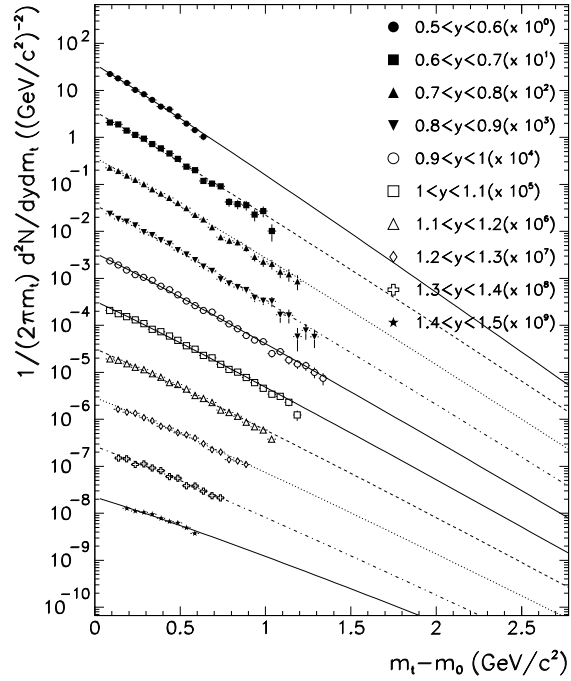,height=4.in}}
\caption{Proton 
$m_t$-spectra for central ($<$5\%) Au-Au collisions at a 
beam kinetic energy of 10.8 GeV/nucleon for different
rapidity bins (laboratory system). Curves are Boltzmann distributions fitted
to the data. Spectra for adjacent rapidity bins are offset by factors of ten
to avoid overlapping.}
\label{Back_fig2}
\end{figure}

\section{Results}

In Fig. \ref{Back_fig2}, spectra of the invariant probability for proton emission per
trigger event are plotted for the 10.8 GeV/nucleon beam energy
as a function of the transverse mass $m_t-m_0$ for the 5\% most central
collisions as determined from the energy deposition in the
zero-degree calorimeter. The spectra are shown for different rapidity bins
as indicated. Only statistical error bars are shown.
Note that adjacent spectra are offset by factors of ten
to avoid overlapping. The range in $m_t$ reflects the acceptance of the
spectrometer in the different rapidity bins ranging from backwards to near
mid-rapidity at $y_{pp}$ = 1.613. The curves represent the best fits to the
spectra using a Boltzmann distribution, {\it i.e.}
\begin{equation}
\frac{1}{2\pi m_t}\frac{d^2N}{dydm_t} = Cm_t\exp(-m_t/T),
\end{equation}
where $C$ is a normalization constant and $T$ is the inverse slope of the
spectrum, both of which are determined from the fit to the data. We observe
that the experimental $m_t$-spectra are in excellent agreement with this
shape although they could also be described almost equally well by a pure
exponential function.

\begin{figure}[ht]
\centerline{
\epsfig{file=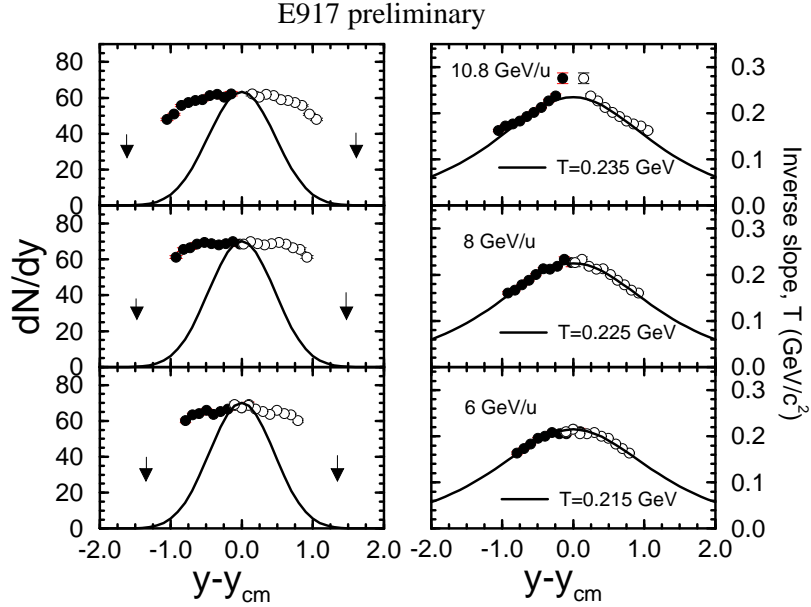,height=4.in}}
\caption{Proton 
rapidity distributions in the center-of-mass system (left
panels) and inverse slopes (right panels) are compared to a simple thermal
source prediction for the three beam energies. The arrows indicate target and beam
rapidities.}
\label{Back_fig3}
\end{figure}

From these fits we derive the total probability for proton emission per unit
of rapidity, $dN/dy$, which is plotted as a function of rapidity in the
center-of-mass frame, $y-y_{cm}$, in the left panels of Fig. \ref{Back_fig3}. The
derived inverse slopes $T$ are shown in the right hand panels. Data are
shown for 5\% central collisions at all three beam energies, where the
centrality for the 6 and 8 GeV/nucleon data are obtained from the
the multiplicity array at the target position. The measured
points are represented by solid circles, whereas reflection around
mid-rapidity results in the open points. Error bars on the fit parameters
are purely statistical and do not include possible systematic errors.

We note that all three $dN/dy$ distributions are quite flat over the
measured rapidity range. The inverse slopes, $T$, show, however, a distinct
peaking at mid-rapidity. The solid curves in Fig. \ref{Back_fig3}
represent the
expectation for isotropic emission from a thermal source with temperature,
$T_0$, at rest in the center-of-mass system.
For such a source one expects that the $y$-dependence of the inverse slope is:
\begin{equation}
T = T_0/\cosh y 
\end{equation}
and a $dN/dy$ distribution of \cite{Schnedermann93}
\begin{equation}
\frac{dN}{dy} \propto
T_0\left(m_0^2+2m_0\frac{T_0}{\cosh y }+2\frac{T_0^2}{\cosh^2 y
}\right)\exp(-m_0\cosh y/T_0),
\end{equation}
where $m_0$ is the proton rest mass. 
In the right hand panels of Fig. \ref{Back_fig3} 
we compare the rapidity dependence of the
inverse slope, $T$, with those predicted by this model. The source
temperature, $T_0$, was adjusted to account for the observed inverse slope at
mid-rapidity. We note that this naive model gives a  rather good 
representation of the observed inverse slopes.

On the other hand, a comparison of the predicted distribution in rapidity
$dN/dy$ with the measurements (left hand panels in Fig. \ref{Back_fig3}) 
reveals a
discrepancy which clearly demonstrates that the observed proton spectra
are inconsistent with isotropic emission from a single source at rest in the
center-of-mass system. Rather, we note that the rapidity distributions for
protons are flat over a  wide range of rapidities
indicating a significant degree of either incomplete stopping or longitudinal
expansion at all three beam energies.

Of course, the  naive model shown here also disregards the possible effects of
radial expansion in the fireball. It has been shown\cite{Schnedermann93},
however, that, within a
wide range of parameters, there is a strong anti-correlation 
between the radial
expansion velocity, $v_0$, and the source temperature,
$T_{source}$, such that
it is impossible to disentangle their relative values from fits to spectra
of a single particle species {\it e.g.} protons. In the present analysis we
have therefore chosen to use only a single parameter, namely the {\it apparent}
source temperature, $T_0$, keeping in mind that 
its value does not necessarily represent the
true temperature of the source formed in a central Au-Au collision.

\begin{figure}[hb]
\centerline{
\epsfig{file=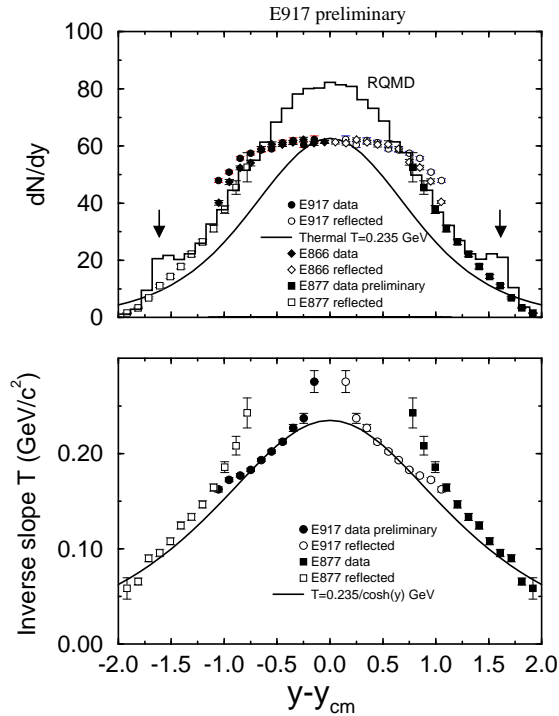,height=4.in}}
\caption{Proton rapidity distributions (top) and inverse slopes (bottom)
for Au-Au collisions at 10.8 GeV/nucleon beam kinetic energy 
compared with previously
published results from the E866 and E877 experiments at the AGS. The arrows
indicate target and beam rapidities.}
\label{Back_fig4}
\end{figure}

In Fig. \ref{Back_fig4} we compare the 
$dN/dy$ (top panel) and inverse slopes $T$ (bottom
panel) from our experiment to
results from the E866 \cite{Ahle98} and E877\cite{Lacasse96}
experiments at the AGS. At
mid-rapidity there is good agreement between
the present data for the $dN/dy$ distribution and that from the
E866 experiment, but a relatively small
discrepancy between the two data sets is apparent
at less central rapidities. The source of this discrepancy is presently
being investigated. The E877 data were measured in the extreme forward rapidity
region but overlap with the present (reflected) data in the rapidity 
region $y-y_{cm}$ = 0.8 - 1.1. Here, there appear to be
substantial discrepancies between the data sets.
The general trend of the three data sets is, however, clear.
There is a wide range -0.7$ < y-y_{cm} <$ 0.7 around mid-rapidity, where
the $dn/dy$ distribution is essentially flat, followed by a monotonic
decrease on either side. We note that this observed range of essentially
constant $dN/dy$ is inconsistent with thermal emission from 
a stopped source (solid curve), as well as early predictions of the 
Relativistic Quantum Molecular Dynamics model\cite{Sorge89} (RQMD v1.08, 
\cite{Lacasse96}) (solid histogram), both of which are too sharply peaked at
mid-rapidity.

The slopes, however, appear to be in reasonable agreement with the $1/\cosh y$
dependence expected from the thermal model (solid curve) although this may
be fortuitous since the rapidity distribution clearly show that there
is either a significant amount of transparency or longitudinal expansion
at these energies which violates
the assumption of a thermal source at rest in the center-of-mass system.

\section{Conclusion}

An analysis of $m_t$-spectra for protons emitted in central Au-Au
collisions at beam kinetic energies of
6, 8 and 10.8 GeV/nucleon in terms of Boltzmann distributions has been
carried out, and the resulting $dN/dy$-distributions and inverse slopes
derived. They are compared to a simple thermal model assuming
isotropic emission from a source at rest in the center-of-mass system
corresponding to complete stopping of the colliding Au nuclei in central
collisions. We find that the rapidity distributions $dN/dy$ are substantially
wider and essentially constant around mid-rapidity 
although the inverse slopes exhibit
the expected $1/\cosh y$ dependence on rapidity. We interpret this as a
manifestation of incomplete stopping or longitudinal expansion 
of the entrance channel momenta at all three beam energies.

\begin{acknowledgments}
This work was supported by the Department of Energy,
 the National Science Foundation, (USA), 
and KOSEF (Korea).
\end{acknowledgments}


\begin{chapthebibliography}{1}

\bibitem{Holzman}
Holzman, B. {\it et al.} (1999) Contribution to these proceedings.

\bibitem{Chang}
Chang, W.-C. {\it et al.} (1999) Contribution to these proceedings.

\bibitem{Schnedermann93}
Schnedermann, E. {\it et al.} (1993) Phys.\ Rev.\ {\bf C48}, 2462.

\bibitem{Ahle98}
Ahle, L. {\it et al.} (1999) Phys.\ Rev.\ {\bf C57}, R466.

\bibitem{Lacasse96}
Lacasse, R. {\it et al.} (1996) Nucl.\ Phys.\ {\bf A610}, 153c.

\bibitem{Sorge89}
Sorge, H. {\it et al.} (1989) Ann.\ Phys.\ (NY)\ {\bf 192}, 266.

\end{chapthebibliography}
\end{document}